# On the distribution of protostar masses

**Philip C. Myers**

Harvard-Smithsonian Center for Astrophysics, 60 Garden Street, Cambridge MA 02138 USA

pmyers@cfa.harvard.edu

**Abstract.** The distribution of protostar masses is studied for core-environment systems whose duration of infall follows a waiting-time distribution. Each core-environment system has a continuous density profile with no barrier to mass flow. The core is an isothermal sphere and the environment is a filament, a layer, or a uniform medium. The infall is terminated by gas dispersal due to outflows and turbulence. The distribution of infall durations is a declining exponential, the simplest waiting-time distribution. The resulting distribution of protostar masses closely resembles the initial mass function, provided the environment density is sufficiently high, and the distribution of initial core masses is sufficiently narrow. The high-mass tail of the mass function increases strongly with environment density and weakly with environment dimension. "Isolated" regions of low environment density form protostars of low mass from within the parent core. In contrast, "clustered" regions of high environment density form protostars of low mass from core gas, and protostars of high mass from core and environment gas.

# 1. Introduction

The initial mass function (IMF) of stars, or the distribution of stellar masses at birth, is a property which apppears similar over a wide range of settings in the Milky Way and in other galaxies (Chabrier 2005). Accounting for the distribution of stellar masses is a key test for any model of star formation (Miller & Scalo 1979, McKee & Ostriker 2007).

It is widely accepted that most stars form in concentrations of dense gas (Beichman et al 1986), and that such young stars are most frequently found in groups and clusters in molecular clouds (Lada & Lada 2003). It is less clear how the mass of a protostar is related to the mass of the dense core where it forms.

In one view, a core is essentially a fixed-mass reservoir of gas, which contributes a significant fraction of its initial mass to its protostar. Then core and protostar masses are proportional, and their mass distributions have the same shape (Motte, André, & Neri 1998, Alves, Lada & Lada 2007). Many authors have suggested ways to form a mass distribution of cores which resembles the IMF, particularly from processes of turbulent fragmentation (Hennebelle & Chabrier 2008 and references therein).

In another view, a core is the densest part of a more extended distribution of gas, with no physical barrier to accretion (Shu 1977, Myers & Fuller 1992, Caselli & Myers 1995, McKee & Tan 2003). Protostars originate in cores, but their masses do not correlate (Bonnell et al 1997, Bate & Bonnell 2005), or their correlation depends on fragmentation and core definition (Swift & Williams 2008), or on the range of gas dispersal times (Myers 2008, hereafter Paper 1). Alternately, their correlation may be coincidental rather than genetic (Hatchell & Fuller 2008).

In Paper 1, protostar masses arise from the competition between infall and gas "dispersal." Infalling gas continues to accrete onto the protostar until it is dispersed due to several possible causes. If the dispersal has a characteristic time since the start of infall, the protostar mass depends on the initial configuration of dense gas and on the dispersal time.

A typical dispersal time is of order 0.1 Myr, according to the statistical frequency of association between protostars and dense cores in the nearby Perseus and Ophiuchus star-forming clouds (Jørgensen et al 2008). In this study, young stellar objects (YSOs) detected by the *Spitzer Space Telescope* are associated with submillimeter wavelength dense cores with decreasing frequency as the evolutionary stage of the YSO increases. The fraction of YSOs with associated cores decreases from 93% for "Class 0" embedded sources to 55% for "Class I" YSOs. This decrease suggests that associated cores generally become undetectable in a time between the typical age of the Class 0 and Class I periods. For constant YSO birthrate, the typical age is half the duration of the period, or 0.05 to 0.2 Myr, according to the "c2d" study of YSOs in the same clouds (Evans et al 2009).

Many factors are expected to remove gas which could otherwise contribute to the final protostar mass. Some dispersal is due to "stellar feedback," including outflows from within the core or from neighboring protostars, and from stellar heating and ionization. Other agents of dispersal include turbulence, competition against nearby accretors, and ballistic motion of the protostar from dense gas to less dense gas, as described in Paper 1.

The present paper studies the distribution of protostar masses arising from an ensemble of accreting "core-environment" systems, on the assumption that the dispersal time has a distribution of values over the ensemble. The main finding is that the protostar mass function resembles the initial mass function of stars (IMF), if the dispersal times have a simple waiting-time distribution, if the initial core masses have a distribution narrower than the IMF, and if the environment gas is as dense as in some embedded clusters.

Section 2 presents relations among density, mass, and free-fall time for core-environment systems. Section 3 computes protostar mass functions for differing environment density and geometry, and compares them to each other and to the IMF. Section 4 discusses the results, and Section 5 gives a summary of the paper. Calculations of mass and free-fall times for nonuniform environment geometry are given in the Appendix.

2. **Core-environment models**

The basic model has a radial density profile which decreases steeply from a central maximum and makes a smooth transition to a profile which is shallower in at least one dimension. This "core-environment" model allows a more realistic descriptions of accreting, star-forming gas than does a model of an isolated core. As in Paper 1, the density is the sum of a core component and an environment component. The core component is a singular isothermal sphere (SIS; Chandrasekhar 1939, Shu 1977). If the initial core component is an isothermal sphere instead of a singular isothermal sphere, the results change slightly at masses below ~0.01 $M_O$, as described in Section 4.3.

More detailed descriptions of cores and their environs are given in Goodman et al (1998), Johnstone, Di Francesco & Kirk (2004), Bergin & Tafalla (2007), Di Francesco et al (2007), Ward-Thompson et al (2007), and Hatchell & Fuller (2008).

*2.1. Densities*

The density of the core-environment system is given by

$$n = n_C + n_E \qquad (1)$$

where the density of the core component is that of a SIS,

$$n_C = \frac{\sigma^2}{2\pi m G r^2} \qquad (2)$$

where $r$ is the radius from the center of the core, $\sigma$ is the 1D velocity dispersion, $m$ is the mean molecular mass, and $G$ is the gravitational constant.

The core and environment components are continuous density functions defined at all radii. At small radius the mass enclosed by the core component dominates the total mass, while at large radius the mass enclosed by the environment component dominates the total mass.

Unlike the pressure-truncated isothermal sphere (Bonnor 1956, Ebert 1955), there is no bounding surface which includes all of the gas available for accretion and excludes all the rest.

The simplest environment component is a uniform 3D medium of density $n_u$,

$$n_E = n_u. \qquad (3)$$

This uniform environment might approximate a self-gravitating region which is warmer and/or more turbulent than the core itself, within its first scale height.

The density profile for a core-environment system with a uniform environment is shown in Figure 1. The density varies continuously with radius, and has no barrier to accretion. Such a system is in the early stages of collapse, since it is the sum of an equilibrium component (the SIS) and a uniform component which adds mass but adds no pressure gradient. The profile resembles an "overdense" initial state, where the mass of an equilibrium structure is incremented in order to initiate its collapse (e.g. Foster & Chevalier 1993, Ogino et al 1999).

Equations for the density profiles of other environmental models, a stratified 2D layer and a centrally concentrated 1D filament, are given in the Appendix.

*2.2. Masses*

For each of the three core-environment models, the mass within spherical radius r is computed by integrating the density profile. When the environment component is uniform this mass is

$$M_u(r) = \frac{2\sigma^2 r}{G}\left(1 + \xi^2\right). \qquad (4)$$

Here $\xi \equiv r/r_0$ is defined in terms of the "equal mass radius" $r_0$ which encloses equal mass in the core and environment components,

$$r_0 \equiv \sigma \left( \frac{3}{2\pi G m n_u} \right)^{1/2}. \qquad (5)$$

The mass of the core component within this radius is called the "core mass" $M_{core}$, given by

$$M_{core} = \left( \frac{6}{\pi} \right)^{1/2} \frac{\sigma^3}{G^{3/2} m^{1/2} n_u^{1/2}}. \qquad (6)$$

Thus the core mass is half the mass of the core-environment system within $r_0$. The core component contains most of the mass within $r$ if $r < r_0$, while the environment component contains most of the mass within $r$ if $r > r_0$. Figures 1 and 2 are marked to show these two regimes. Similar expressions for core-environment mass are given in the Appendix for environments which are layers and filaments.

This definition of core mass is analogous to some observational practice, where a background component is subtracted from the mass within a radius which marks the transition from steep to shallow slope. In relation to the Jeans mass $M_J$ of the uniform medium (Spitzer 1978), $M_{core}/M_J = \sqrt{6}/\pi^2 \approx 1/4$. Note that the "core mass" is a fixed quantity for a given temperature and environment density, while the "mass of the core component" is an increasing function of radius.

Figure 2 shows the dependence of core-environment mass on radius $M(r)$ for the three environments considered. For each core-environment model, $M(r)$ increases with $r$ more rapidly than for the core alone. Also, $M(r)$ increases more rapidly when the environment component is a uniform medium than when it is a layer, and slightly more rapidly when the environment component is a layer than when it is a filament, for the same temperature and peak environment density.

As in Paper 1, it is useful to express $M(r)$ in terms of the free-fall time. The free-fall time $t_f$ for pressure-free, spherical infall is obtained from

$$t_f^2 = \frac{\pi^2 r^3}{8GM(r)} \qquad (7)$$

(Hunter 1962, Spitzer 1978). For the core-environment system with uniform environment, this mass is given by

$$M = M_{core}\theta\left(1-\theta^2\right)^{-3/2} \qquad (8)$$

where the dimensionless time $\theta$ is the ratio of the free-fall time $t_f$ of the core-environment system to the free-fall time $t_u$ of the uniform environment alone,

$$\theta \equiv \frac{t_f}{t_u} \qquad (9)$$

with $0 \leq \theta < 1$.

The mass available to accrete onto a protostar increases with free-fall time, as shown in Figure 3 for isolated cores and for environments which are filaments, layers, and uniform media. Each environment has the same peak density. For core-environment systems, the available mass increases linearly at early times, and more rapidly at later times. The mass of a core-environment system increases faster with $t_f$ than for a core alone. Also, the mass of a core-environment system increases faster with $t_f$ when the environment component is uniform than when it is a layer, and slightly faster when the environment component is a layer than when it is a filament.

*2.3. "Isolated" and "clustered" masses*

The mass available to accrete onto a protostar in a free-fall time increases with environment density. In turn, this density can discriminate between "isolated" and "clustered" conditions in star-forming regions. To illustrate these points the "equal-mass" environment density $n_{E0}$ is defined in analogy with the equal-mass radius in equation (5):

$$n_{E0} \equiv \frac{3\pi\left(1 - 2^{-2/3}\right)}{32 m G t_f^2}. \qquad (10)$$

In equation (10), $n_{E0}$ is the environment density whose associated core-environment mass is twice the mass of the core component having the same free-fall time. Then the core component has most of the core-environment mass when $n_E < n_{E0}$, while the environment component has most of the core-environment mass when $n_E > n_{E0}$. For $t_f = 0.1$ Myr, equation (10) gives $n_{E0} = 4 \times 10^4$ cm$^{-3}$.

The mass available in a free-fall time increases strongly with environment density. The available mass has been calculated as a function of $n_E$, for the same systems as in Figures 2 and 3, with fixed temperature and free-fall time. Figure 4 shows that the available mass has two distinct regimes, depending on the value of $n_E$. When $n_E < n_{E0}$ the available mass is low, nearly constant, and most of this mass is due to the core component. In contrast, when $n_E > n_{E0}$ the available mass is higher, increases with $n_E$, and most of the mass is due to the environment component. An order of magnitude increase in $n_E$ can give an order of magnitude increase in the mass available in a free-fall time.

The environment density which discriminates mass available in a free-fall time also discriminates regions of "isolated" and "clustered" star formation. In regions of isolated star formation, low-mass stars form in widely separated cores with low intercore extinction. The typical spacing of young stellar objects (YSOs) in the filamentary regions of the Taurus complex is 0.3 pc, matching the Jeans length for 10 K gas and density $3 \times 10^3$ cm$^{-3}$ (Spitzer 1978, Hartmann 2002). The gas between cores has a few magnitudes of visual extinction, corresponding to column density of a few $10^{21}$ cm$^{-2}$ (Dobashi et al 2005).

In clustered conditions both low-mass and massive stars form, with closer spacings and greater intercore extinction (Herbig 1962, Lada, Strom & Myers 1993, Ward-Thompson et al 2007, Allen et al 2007, Megeath, Li & Nordlund 2009). For temperature 10 K and for environment density $8 \times 10^4$ cm$^{-3}$ the Jeans length is 0.075 pc, matching the median spacing of YSOs in 39 nearby embedded clusters, which have mean visual extinction exceeding 8 magnitudes (Gutermuth et al 2009).

The environment density which separates the isolated and clustered regimes in Figure 4 can also be expressed in terms of column density, if the environment is an isothermal filament or layer. The equal-mass column density along a line normal to the axis of the isothermal filament, or along a line normal to the midplane of the isothermal layer, is of order $N_{E0} \sim \sigma/(mGt_f)$. For temperature $T = 10$ K and free-fall time 0.1 Myr, $N_{E0} = 0.8 \times 10^{22}$ cm$^{-2}$ for the layer and $N_{E0} = 1.0 \times 10^{22}$ cm$^{-2}$ for the filament. If the gas temperature exceeds 10 K, the critical column density increases as $T^{1/2}$.

Thus when intercore column densities in a star-forming region exceed $\sim 10^{22}$ cm$^{-2}$, this core-environment model indicates that there is sufficient gas available to form massive stars in a clustered setting, in times of order 0.1 Myr. It is notable that the nearest young stellar groups with at least 20 members have mean column density greater than $1 \times 10^{22}$ cm$^{-2}$ (Gutermuth et al 2009, Myers 2009).

*2.4. Accretion*

The accreted mass is here called the "protostar mass" $M_\star$, or the mass of the protostar-disk system. This protostar mass equals the foregoing mass available in a free-fall time, if the infall is sufficiently cold and spherical, if all mass shells start collapsing from rest at the same time, and if the initial mean density decreases monotonically with radius so that collapsing shells do not cross (Hunter 1962, Spitzer 1978). Under more realistic conditions, the accretion flow is expected to be pressurized (Bondi 1952), nonsteady (Vorobyov & Basu 2005), and geometrically complex, reflecting the infall onto the accretion disk (Terebey, Shu & Cassen 1984). Further the mass available in a free-fall time decreases if the accretion proceeds in "inside-out" fashion (Shu 1977) and increases as the initial infall velocity increases from zero (Fatuzzo, Adams & Myers 2004).

Therefore the protostar mass arising from a core-environment system whose environment component is uniform is based on equation (8) for the mass available in a free-fall time, according to

$$M_\star = \varepsilon M_{core} \theta \left(1 - \theta^2\right)^{-3/2} \qquad (11)$$

where $\theta$ is defined as in equation (9), with $t_f$ set equal to the time $t$ since the start of collapse. The "accretion efficiency" $\varepsilon \equiv M_\star/M$ is a constant parameter representing the departure of the actual accretion flow from the cold, steady, spherical flow assumed in the present model. This efficiency should not be confused with the ratio of protostar mass to core mass, which is often called the "star formation efficiency."

If the environment density $n_E$ is fixed, equation (11) and Figure 3 indicate that $M_\star$ increases approximately linearly with time for early times and more steeply for late times. At early times when $\theta \ll 1$, the mass accretion rate $dM_\star/dt$ approaches

$$\left(\frac{dM_\star}{dt}\right)_{early} = \frac{8\sigma^3\varepsilon}{\pi G} . \qquad (12)$$

This constant accretion rate is greater by a factor $8\varepsilon/(0.975\pi)$ than for inside-out isothermal collapse (Shu 1977). Equivalently, the accretion efficiency of inside-out collapse is $\varepsilon = 0.383 = 0.975\pi/8$.

At late times when $1-\theta << 1$, the accretion rate approaches

$$\left(\frac{dM_\star}{dt}\right)_{late} = \frac{24\sigma^3\varepsilon}{\pi G}\left(\frac{M_\star}{M_{core}}\right)^{5/3} \qquad (13)$$

using equations (7) and (11). This late-time accretion rate is significantly greater than at early times. Its power-law dependence on $M_\star$ is similar to that of Bondi accretion, whose accretion rate is proportional to $M_\star^2$ (Bondi 1952, Zinnecker 1982, Shu 1992). For example when $\theta=0.8$, $M_\star/M_{core}=3.7\varepsilon$. Then $dM_\star/dt$ is less than the Bondi rate by a factor 5.4 and greater than the early-time rate by a factor 27.

This rapid increase in $dM_\star/dt$ from low to high protostar mass accords with the view that massive protostars form in a small fraction of an embedded cluster lifetime. Thus a protostar of mass 10 $M_O$ which forms in ~0.1 Myr has a mean mass accretion rate ~$10^{-4}$ $M_O$ yr$^{-1}$, matching an estimate based on studies of outflows (Zhang et al 2005). This rate exceeds that estimated for isolated low-mass stars at 10 K (Shu 1977) by a factor ~50.

These results indicate that core environment and the termination of accretion can play a major role in setting the protostar mass. The next section gives a more quantitative description of protostar mass distributions.

## 3. Mass functions

The relationship of the protostar mass function to initial gas properties is an important aspect of star formation studies. However a consensus has not yet been reached about this relationship.

In one well-known picture, protostars and cores have similar mass functions, because a core can be considered to be distinct from its surrounding gas. The efficiency of converting core mass into protostar mass is < 1 because gas does not accrete from beyond the core. This picture has observational support, since mass functions of observed cores have been found to be similar to the IMF in numerous studies (Motte, André, & Neri 1998, Alves, Lada & Lada 2007, Ward-Thompson et al 2007, Rathborne et al 2009).

On the other hand, the shape of the mass function is expected to depend on how much one core varies from the next in its fragmentation, its star formation efficiency, its evolutionary timescale, and its mass available for accretion (Hatchell & Fuller 2008, Goodwin et al 2008, Swift & Williams 2008). In this view, similarity between observed core mass functions and the IMF may be coincidental (Hatchell & Fuller 2008).

Paper 1 noted that the similarity of core and protostar mass functions is expected only if the dispersal time scale has a relatively small variation from core to core. Similarly, equation (11) indicates that a constant ratio of $M_*/M_{core}$ requires $\theta$, the ratio of the duration of accretion to the environment free-fall time, to be limited to a single constant value.

This section investigates the alternative possibility that accretion may endure over a range of times, and presents mass functions arising from a simple distribution of accretion time.

### 3.1. Dispersal and accretion times

In Paper 1, the final protostar mass for a core with a uniform environment was approximated as the initial gas mass whose free-fall time is equal to a characteristic dispersal time scale $t_d$, provided $t_d$ is not too close to the free-fall time $t_u$ of the environment alone. In this paper, the final protostar mass is again estimated from the free-fall accretion of the initial gas, which endures until a time characteristic of the core-environment gas dispersal.

It is further assumed that this dispersal time varies from one core-environment system to the next, because of the many ways to disperse dense gas. Outflows are the best-understood mechanism of dispersing a parent core (Arce et al 2007). However, dense gas can also be dispersed by outflows from nearby cores, by internal and external heating, by turbulence and ionization, by competition for mass among nearby accretors, and by migration of YSOs away from their original dense gas, as discussed in Paper 1. The effect of an external outflow may extend beyond the time when the outflow is detectable, since the low-density cavity of a "fossil" outflow may continue to expand long after the passage of the initial protostellar jet (Quillen et al 2005, Cunningham et al 2006).

The numerous mechanisms of dispersal suggest that protostar accretion may endure for a significant range of times. Further, no physical reason has been suggested why accretion should last until only one particular time, as opposed to enduring over a range of times. Thus the distribution of durations is assumed here to have nonzero width, as was also assumed by Myers (2000) and by Basu & Jones (2004) for accretion onto cores.

Henceforth "dispersal time," "accretion time" and the "waiting time" until accretion is negligible have the same meaning and will be used interchangeably.

The allowed range of accretion times is assumed to have a lower limit of zero, since dense gas can become unbound before it can form a protostar. Such unbinding could arise from external turbulence or outflows dispersing a condensation which is only lightly bound. For the uniform environment, the range of accretion times also has a finite upper limit, given by the free-fall time of the uniform medium alone. Note that this upper limit will increase if the environment gas has additional means of support against collapse, such as an inward gradient of magnetic pressure.

*3.2. Distribution of accretion times*

The distribution of accretion times in star-forming regions is unknown, and the first empirical estimates may come from statistical studies of association of protostars and cores in nearby star-forming complexes. Here a simple distribution is used as an example, which may guide more realistic studies in the future.

The basic idea is that infalling dense gas may be dispersed at any time before it reaches the protostar, terminating the accretion. Then the likelihood that this gas does not disperse until a particular time decreases with time. Specifically, it is assumed that dispersal in an interval of fixed duration is equally likely, independent of when the interval occurs. The time until accretion stops then has a "waiting-time" distribution of probability density (Feller 1968, Nadarajah 2007).

When the core environment is uniform, the waiting time has a finite maximum, equal to the uniform free-fall time. This distribution is said to have a "finite time horizon" (Barlow 1998). If this maximum waiting time is much greater than the mean waiting time, the resulting distribution approaches the exponential distribution

$$p(\theta) = \frac{1}{\bar{\theta}}\exp\left(-\frac{\theta}{\bar{\theta}}\right). \tag{14}$$

This distribution also arises from the Poisson approximation to the binomial distribution (Feller 1968, Basu & Jones 2004).

In equation (14), $\theta = t_w/t_u$ is the waiting time normalized by the free-fall time of the uniform environment gas. Since the waiting time is essentially the duration of free-fall accretion until dispersal, $\theta$ has the same definition as in equations (9) and (11). Similarly, $\bar{\theta} = \bar{t}_w/t_u$ is the normalized value of the mean waiting time.

### 3.3. Mass functions for identical cores

The protostar mass function depends on the distribution of accretion times, discussed above, and on the distribution of initial core masses. For simplicity, protostar mass functions are first discussed for identical core-environment initial states. Section 3.4 discusses mass functions arising when both waiting times and core masses are distributed.

The identical core-environment systems considered here have the same temperature $T$, and their environment components are uniform, with the same value of $n_E$. Then from equations (6) and (11), $\varepsilon M_{core}$ is constant and the mass function depends on only one additional parameter, the mean waiting time $\bar{t}_w$ or equivalently its normalized value $\bar{\theta}$.

The mass function of protostars is denoted $\Phi$, and is defined as the number of protostars per logarithmic mass interval, or

$$\Phi \equiv M_\star \, p(M_\star). \tag{15}$$

Here $p(M_\star)$ is the probability density that the final protostar mass lies between $M_\star$ and $M_\star + dM_\star$. Since $M_\star$ increases monotonically with $\theta$ as in equation (11), when $\varepsilon M_{core}$ is fixed the mass function can be expressed in terms of $\theta$ as

$$\Phi = M_\star \, \frac{p(\theta)}{dM_\star/d\theta}. \tag{16}$$

Using equation (11) to evaluate the derivative in equation (16), along with $p(\theta)$ from equation (14), gives the identical-core mass function in terms of $\theta$ and its mean value $\bar{\theta}$,

$$\Phi = \frac{1-\theta^2}{1+2\theta^2} \frac{\theta}{\bar{\theta}} \exp\left(-\frac{\theta}{\bar{\theta}}\right). \tag{17}$$

When this function is evaluated in terms of protostar mass, its shape depends only on the parameter $\bar{\theta}$ and its position is set by the modal mass $M_{\star m}$. Fitting trials indicate that the shape and modal mass have combined best match to those of the IMF (Kroupa 2002) when $\bar{\theta} = 0.20$ and $M_{\star m} = 0.16\ M_O$. Figure 5 shows the best-fit MF with these properties. This model matches the position and asymmetric shape of the IMF fairly well, with only two independent parameters.

The shape and modal mass of the mass function are related to other model parameters. Differentiating the mass function gives the modal mass in the limit $(\bar{\theta})^2 \ll 1$ as

$$M_{\star m} \approx \frac{8\sigma^3 \bar{t}_w \varepsilon}{\pi G}. \tag{18}$$

Then the environment density and mean waiting time are related by

$$\left(\frac{n_E}{10^4\ \text{cm}^{-3}}\right) = \left(\frac{\bar{\theta}}{0.20}\right)^2 \left(\frac{\bar{t}_w}{0.07\ \text{Myr}}\right)^{-2} \tag{19}$$

based on the definition of $\bar{\theta}$. The accretion efficiency is related to the temperature and environment density by

$$\varepsilon = 0.58 \left(\frac{M_{\star m}}{0.16\ M_O}\right) \left(\frac{\bar{\theta}}{0.20}\right)^{-1} \left(\frac{n_E}{10^4\ \text{cm}^{-3}}\right)^{1/2} \left(\frac{T}{10\ \text{K}}\right)^{-3/2} \tag{20}$$

based on equations (6), (18) and (19).

Equations (19) and (20) indicate various ways for the initial density and temperature to be consistent with the mass function in Figure 5. Examples of consistent cases include (a) cold isolated gas, with $n_E = 1 \times 10^3$ cm$^{-3}$ and $T = 10$ K, implying $\bar{t}_w = 0.22$ Myr and $\varepsilon = 0.18$; (b) cold clustered gas, with $n_E = 3 \times 10^4$ cm$^{-3}$ and $T = 10$ K, implying $\bar{t}_w = 0.04$ Myr and $\varepsilon = 1$; and (c) warm clustered gas, with $n_E = 3 \times 10^4$ cm$^{-3}$ and $T = 20$ K, implying $\bar{t}_w = 0.04$ Myr and $\varepsilon = 0.35$.

For massive star formation, the low-density isolated case (a) is unrealistic because the infall must be too extended in both space and time. A 10 M$_O$ protostar would require gas from within an initial radius 0.58 pc, comparable to the radius of an entire cluster. This gas would take 0.97 Myr to accrete, comparable to the star-forming life of a cluster.

In contrast, the high-density clustered cases (b) and (c) are more realistic. The gas which supplies a 10 M$_O$ protostar has initial radii 0.15 and 0.10 pc, and their accretion times are 0.18 Myr in each case. These initial radii lie within the typical radius of an embedded cluster, and this accretion time lies within the typical star-forming duration of an embedded cluster. The corresponding mean mass accretion rate lies within a factor 2 of $\sim 10^{-4}$ M$_O$ yr$^{-1}$, a characteristic value for massive protostars (Zhang et al 2005).

*3.3.1. Varying environment density* Mass functions are computed for identical cores with a uniform environment having "clustered" and "isolated" values of density. The parameters are those of example (b) above, except that the isolated environment density is decreased tenfold to $3 \times 10^3$ cm$^{-3}$. Figure 6 shows that the resulting mass functions are nearly identical at low mass, due to accretion from core gas. The high-mass tail appears only for the clustered environment, due mainly to accretion from the environment component. These properties are expected from the difference in available mass in a free-fall time between clustered and isolated densities, shown in Figure 4.

*3.3.2. Varying environment geometry*  The similarity of the protostar mass function to the IMF for uniform initial environments is also seen for nonuniform initial environments. Mass functions are calculated for isolated cores and for core-environment systems whose environments are filaments, layers, and a uniform medium.  All systems have the same waiting-time distribution of accretion times, the environments have the same peak density, and the mass functions are calculated in the same way from equation (16), assuming the parameters listed in case (b) above.

The resulting mass functions are shown in Figure 7.  They have essentially the same properties as the mass functions having high and low environment density, in Figure 6.  The mass functions vary more with environment density than with environment dimension. Comparing their high-mass tails, the "maximum mass" where the mass function is 1% of its peak value increases as the environment dimension increases from 1D filaments to 2D layers to the 3D uniform medium.  This sequence is expected from the similar ranking of mass available in a free-fall time in Figure 3.

*3.3.3. Varying accretion efficiency, temperature and mean waiting time*  Mass functions were also calculated for cores with a uniform environment having various values of $\varepsilon$, $T$, and $\bar{t}_w$.  As equation (11) indicates, the modal mass increases as $\varepsilon^1$ and as $T^{3/2}$ with no change in the shape of the mass function.  Increasing the mean waiting time $\bar{t}_w$ increases the modal mass as $\bar{t}_w^{\,1}$ and also increases the high-mass tail, as the likelihood of accretion from the environment increases.

*3.3.4. No massive protostars from environments of low density and dimension*  The dependence of the mass function on environment density and dimension, shown in Figures 6 and 7, predicts that low-extinction filaments and linear chains of globules are the environments least likely to produce massive protostars.  The main reason is that their low environment density causes their accretion time to exceed their likely gas dispersal time.  The accretion time of a massive protostar increases as environment dimension decreases, as is evident in Figure 3.  In addition, the low dimension of filaments causes their accumulation length to exceed the typical size of a cluster in which massive stars form.  The increase of accumulation length as environment dimension decreases is evident in Figure 2. As a result, mass functions of protostars in low-density filaments should decline more steeply at high mass than mass functions of protostars forming in denser environments of higher dimension.

*3.4. Mass functions for distributed cores*

The identical-core mass function in Figure 5 closely resembles the stellar IMF in its mode and shape, but is somewhat narrower than needed for an exact match. Also, its initial state ignores the range of core masses observed in many studies of star-forming clouds (Motte, Andre, & Neri 1998, Rathborne et al 2009). Thus the core-environment accretion model is combined with an appropriate distribution of core masses, denoted as a "CMF."

It is difficult to combine the core-environment accretion model directly with a CMF obtained from observations. Each model core hosts the formation of one star, while the lowest-mass observed cores are unlikely to make any stars, since they are not gravitationally bound (Enoch et al 2008). Similarly, the highest-mass observed cores are likely to make more than one star, as in B59 in the Pipe Nebula, which hosts more than 10 YSOs (Brooke et al 2007). To compare with the single-star accretion model, an observed CMF should be modified into a "single-star CMF."

Such modification of an observed CMF could proceed by removing the lowest-mass unbound cores and by removing the highest-mass cores which make multiple stars. If these high-mass cores make a total of N stars, they should be replaced in a modified CMF by N lower-mass cores of the same total mass. A related procedure was carried out by Swift & Williams (2008), as discussed in Section 4. However, the number of stars to be formed by an observed core is highly uncertain, and this uncertainty limits the utility of such a modified CMF.

Instead, the CMF is assumed here to have a log-normal form, since simulations of turbulent molecular clouds give distributions of clump mass which are approximately log-normal (Padoan, Jones, & Nordlund 1997, Klessen 2001, Falle & Hartquist 2002). Also, any property which is a product of independent random variables tends to develop a log-normal distribution (Adams & Fatuzzo 1996).

This log-normal distribution of core masses is combined with the waiting-time distribution of accretion times in equation (14), to produce a protostar mass function due to both distributions. As before, the "core mass" is given by equation (6). For the distribution of core masses considered here, the environment density is constant over the distribution. Then the variation of core mass is associated with the variation of only one property, the core temperature.

The mass function is obtained by integrating over the probability distributions (Davenport & Root 1958, Basu & Jones 2004),

$$\Phi = \int_{\theta_{min}}^{\theta_{max}} d\theta\, p(\theta) M_{core}\, p(M_{core}) \qquad (21)$$

where $p(M_{core})$ is given by

$$p(M_{core}) = \frac{1}{\sqrt{2\pi}\, M_{core}\, \sigma_{core}} \exp\left[-\frac{(\ln M_{core} - \mu)^2}{2\sigma_{core}^2}\right]. \qquad (22)$$

Here $p(M_{core})$ is the log-normal probability density that a core has mass between $M_{core}$ and $M_{core} + dM_{core}$, and where $\mu$ and $\sigma_{core}$ are respectively the mean and standard deviation of the distribution of $\ln M_{core}$.

To evaluate equation (21), $M_{core}$ is expressed in terms of $M_\star$ and $\theta$ using equation (11) for the uniform environment. For this environment $\theta_{min} = 0$ and $\theta_{max} = 1$. The mean value $\mu$ is obtained from the parameters used for the identical-core MF in Figure 5, with $T=10$ K and $n_E = 3 \times 10^4$ cm$^{-3}$, giving $M_{core}/M_\odot = e^\mu = 0.80$. The width of the log-normal CMF is $\sigma_{core} = 0.5$. This value was adjusted until the resulting MF matched the IMF of Kroupa (2002).

Figure 8 compares this protostar MF based on distributed cores with that based on identical cores, as in Figure 5, and compares each of these with the IMF. The MF for distributed cores has the same mode and shape as for identical cores, but has slightly greater width and slightly less concave shape, so it matches the IMF more closely.

The log-normal CMF has $\sigma_{core}$ =0.5 and thus has half-maximum (HM) masses 0.44 and 1.4 $M_O$. The ratio of greater and lesser HM core masses is 3.3, smaller by a factor 5.5 than the ratio of greater and lesser HM stellar masses in the IMF of Kroupa (2002). The single-star CMF is much narrower than the IMF, since protostar masses can be less than, or greater than the masses of their parent cores, as discussed earlier.

Similarly, the width of the single-star CMF is narrower than for well-studied observed CMFs, such as in the Pipe Nebula (Rathborne et al 2009). Some such decrease in distribution width is expected because a single-star CMF is narrower than the corresponding observed CMF, as discussed above. Further, the gas temperatures which account for these half-maximum core masses are 7 and 15 K. This range is similar to the range of temperatures derived from $NH_3$ line observations of 70 dense cores in Perseus, 9-15 K (Enoch et al 2008, Rosolowsky et al 2008).

In Figure 8, the combination of the waiting-time distribution and the CMF matches the IMF extremely well, but this match is not unique. For example an equally good match to the IMF results from the "opposite" case where the CMF has the same shape and nearly the same width as the IMF, and where the distribution of infall times is much narrower than assumed in Section 3.3. However as discussed earlier, it is difficult to understand the physical basis of such a narrow distribution of infall times.

It is not possible to match the IMF with a combination of the waiting-time distribution in Section 3.3 and a CMF which has the same shape and width as the IMF, since the combination of these two distributions is broader than either distribution by itself.

## 4. Discussion

*4.1. Assumptions*

In this paper, the basic initial condition is a core-environment system, and all of its mass is available for accretion until it is dispersed. The transition from a core to its environment is therefore not a barrier to mass flow. This idea was assumed and discussed in Paper 1. It reflects the model of Shu (1977), where gas was considered available for accretion with no outer boundary. This initial state differs from the well-known Bonnor-Ebert (BE) sphere (Bonnor 1956, Ebert 1955), where mass available for accretion is limited by a pressure-truncating boundary between a cold dense interior and a hot rarefied exterior. The BE sphere has been invoked in many discussions linking core mass to protostar mass (for a recent review, see Ward-Thompson et al 2007).

Star-forming accretion is assumed to endure for a range of times, due to the variety of ways to disperse dense gas. Thus a given core-environment system can form a protostar whose mass is less than, equal to, or greater than the mass enclosed by the core-environment transition. This idea resembles those in Adams & Fatuzzo (1996), who argued that outflows terminate accretion and have a distribution of properties. The exponentially declining probability of accretion, assumed here, was also used to describe the ejection of protostars from their dense cores due to dynamical interactions (Bate & Bonnell 2005). A similar probability of stopping accretion was assumed in models of growth of dense cores (Myers 2000, Basu & Jones 2004).

*4.2. Predictions for isolated and clustered star formation*

In this paper a simple model of star formation accounts for the shape and modal mass of the IMF in conditions of clustered star formation, and predicts a smaller proportion of massive stars in regions of isolated star formation, where the environment density is an order of magnitude lower.

When the environment density is low, the mass function lacks the high-mass tail seen in clusters, because the accretion time for massive stars exceeds the likely time of gas dispersal. When the environment has both low density and filamentary geometry, massive star formation is even less likely because the necessary accumulation length exceeds the likely extent of a cluster in which massive stars are found. This prediction may be testable in nearby star-forming clouds.

When the environment density is sufficiently high, the model predicts a greater proportion of massive stars than would be expected from the IMF. It may be possible to test this prediction with studies of gas properties and massive star fractions in infrared dark clouds which have temperatures comparable to those in nearby clusters, but greater densities (Jackson et al 2008). A step in this direction has recently been reported (Ragan, Bergin & Gutermuth 2009).

The dependence of the model MF on environment density and geometry does not necessarily violate the "universal" nature of the IMF. Rather, it may indicate that such low-density and high-density environments together contribute too few protostars to alter the IMF of most field stars, which may be born in more intermediate environments. It may be useful to test this idea by studying the distribution of embedded clusters as functions of their number of members and of their mean gas density.

*4.3. Limitations*

In star-forming clouds the forces due to magnetic fields and turbulent motions are not negligible, as discussed in Paper 1. Nonetheless, once a system of cores is established in a star-forming environment, their evolution into protostars has been approximated by many authors solely as a problem in thermal and gravitational physics (Larson 2005). The magnetic and turbulent effects are expected to change results by a factor of order 2, but not to change the qualitative picture. In this paper the "accretion efficiency" $\varepsilon$ summarizes the departure of the protostar mass from that expected in a nonmagnetic, nonturbulent, smooth, steady, spherically symmetric flow. This parameter is found to have values of order 1/2 in order to match the mode of the IMF, for temperatures and mean densities expected in cluster-forming regions. This order-unity value of accretion efficiency tends to corroborate the simple approach adopted here. However, a more serious test requires detailed simulations of systems with infall and dispersal.

The exponential distribution assumed in Section 3 is the simplest nonsingular waiting time distribution. When combined with the core-environment models of section 2, it gives a reasonable approximation to the IMF, as shown in Figure 5. Indeed it can be shown that the identical-core mass function for a uniform environment can achieve a much closer fit to the IMF with only a slight modification of the exponential waiting-time distribution. A more complete understanding requires a more physical explanation of the distribution of accretion times, in terms of the mechanisms which limit accretion. Similarly, there is little observational basis presently available to guide the choice of waiting time distribution. A promising approach for the future may be to estimate the survival probability of dense cores with associated protostars.

The mean waiting time before dispersal is typically $\bar{t}_w$ = 0.04-0.07 Myr in order to provide a good match to the IMF, for the initial temperatures and densities adopted in Section 3. This waiting time is an order of magnitude less than the duration of the "Class I" stage of star formation, 0.54 Myr, according to the "c2d" study of nearby clouds (Evans et al 2009). These times conflict if the mass accretion rate is nearly constant over the entire Class I period. However, many authors have suggested that the mass accretion rate is greatest during the "Class 0" stage when the protostar has a dense circumstellar envelope (André, Ward-Thompson & Barsony 1993, White et al 2007). When the c2d sample is further subdivided into Class 0 and Class I, the Class 0 stage has extinction-corrected duration 0.10 Myr (Evans et al 2009), in better agreement with the mean waiting time obtained here. As with the distribution of waiting times, the mean waiting time may be usefully constrained by more detailed observational estimates.

More generally, it is important to explain why the mean waiting time has an optimal value (about 0.2 of the environment free-fall time), which gives the closest match of the mass function to the IMF. The present model cannot be considered a full explanation of the IMF without a more physical understanding of this condition. In future studies, it may be useful to investigate the idea that protoclusters which satisfy this condition form more stars than do protoclusters having shorter or longer mean waiting times.

The dense core component is modelled in Section 2 as a singular isothermal sphere (SIS), which approximates the isothermal sphere (IS) in the limit of infinite central density (Chandrasekhar 1939). The SIS model is used here because the main focus is on the core environment, and because the SIS gives simpler analytic expressions. If instead the core component is modelled as an IS as in Paper 1, the low-mass part of the identical core mass function differs slightly from its shape based on the SIS. At masses < ~0.01 $M_O$, much lower than the modal mass, the log-log slope approaches 4/3 for the IS instead of 1 for the SIS. After combination with the initial distribution of core masses as in Section 3.4, it seems unlikely that this change could be discerned, within current observational and statistical uncertainties.

*4.4. IMF and CMF*

The origin of the IMF is often considered in two parts: first explaining the distribution of core properties, and then explaining the mass distribution of protostars arising from these cores. This framework is motivated by observed core mass distributions, which frequently follow a high-mass power-law, or a high-mass power-law and a low-mass turnover, with some similarity to the IMF. Thus many authors have studied the first problem, with the goal of producing a core mass distribution matching the shape of the IMF (see Hennebelle & Chabrier 2008 for a review). Implicit in this approach is the idea that the core-environment transition encloses the only gas available for accretion.

In contrast, this paper studies the second part of the problem. It models the formation of protostars from core-environment systems, assuming that the initial system has a continuous density profile, with no accretion barrier. The final protostar mass depends on the duration of accretion, which is likely to vary over a distribution. Thus the distribution of cores which make single stars ("single-star CMF") is one of two distributions responsible for the IMF, and it is narrower than the IMF itself. This single-star CMF is also narrower than the typical mass distribution of observed cores, whose lowest-mass members make no stars and whose most massive members make multiple stars.

The relation of the mass distributions of cores and protostars was studied by Swift & Williams (2008, hereafter SW). SW considered an initial CMF having the same shape as the IMF, whose cores form stars of lower mass according to several prescriptions. For typical statistical uncertainties, SW find that the "overall shape" of the mass function is "robust against different core evolution scenarios." They also find that the position of the peak and the low-mass and high-mass slopes vary for particular prescriptions of core fragmentation.

The SW model assumes that the mass of a protostar is always less than that of its parent core. This requirement tends to shift and stretch the initial distribution toward lower masses. If the typical value of star formation efficiency is comparable to the spread in efficiencies, the main effect is to preserve the shape of the initial distribution, as SW find.

In the present core-environment model, a protostar can be less massive or more massive than the core in which it forms. In this case the protostar mass function is generally broader than the initial core mass distribution and can have different shape, in contrast to the result of SW. For example if the rate of mass increase is proportional to the initial mass, a mass function which is initially log-normal develops a high-mass tail (Basu & Jones 2004).

Hatchell & Fuller (2008) discuss four factors whose variation may cause the protostar mass function to differ in shape from the initial CMF, drawing on results from Goodwin et al (2008) and on SW. Of these four, the factor most relevant to the present work is the ratio of core mass to the mass available for accretion. This ratio varies significantly from core to core because the available mass is the product of the mass available in a free-fall time, the accretion efficiency, and the duration of the infall. The duration of the infall is not constant from core to core, but follows the waiting-time distribution.

*4.5. Low-mass and massive protostars*

In this paper the difference between low-mass and massive protostars is primarily in the duration of their accretion rather than in the initial mass of the cores where they form. Low-mass protostars accrete for a short period, from gas within the core. In contrast, massive protostars accrete for a longer period, from gas within and beyond the core. For a core-environment system with uniform environment, the time to make a star of mass $M_*$ is given by equation (11). For low masses, this time increases linearly with $M_*$, following a mass accretion rate as in equation (12). For high masses, this time approaches the constant value of the free-fall time of the uniform environment, $t_u$.

Thus the star formation time resembles the increasing value of Shu (1977) for low masses and resembles the nearly constant value of Myers & Fuller (1992) for high masses. For massive protostars this nearly constant time differs from that of "competitive accretion models" (e.g.

Bonnell, Bate & Vine 2003), where the protostar gains mass continuously due to moving accretion though the protocluster gas.

The formation of low-mass and high-mass protostars in the same environment resembles that in a recent simulation of a cluster-forming region (Wang et al 2009). There the peak initial gas density in a cluster-forming clump is $3 \times 10^4$ cm$^{-3}$, the same as assumed in cases (b) and (c) of section 3.3. The gas is subjected to turbulent stirring. Both low-mass and massive protostars are formed in the same simulation. The massive protostars are surrounded by "cores" or condensations of ~0.1 pc diameter, but such protostar mass arises from infall motions of the "clump" gas, extending well beyond the core. The accretion of such massive protostars is opposed by winds and outflows from already formed lower-mass stars. As in the present paper, this simulation shows that competition between accretion and dispersal leads to low-mass and massive protostars in a cluster setting, where massive protostars arise from gas far beyond their associated cores.

### 4.6. *Comparison with other models*

The present model differs from the "turbulent core" model of McKee & Tan (2003) in that a massive protostar does not require a "core" which is unusually warm or turbulent in order to form. The core is the starting point for the accretion, but most of the mass arises because the environment beyond the core is unusually dense and/or the waiting time until dispersal is unusually long. For massive protostars this model resembles that of Bonnell, Bate, & Vine (2003) in that most of the mass arises from beyond the immediate vicinity of the protostar. However in this model the accretion differs from that of Bonnell et al (2003) because it is stationary, not moving, and it is limited by dispersal, not by competition with other accretors.

The present model is perhaps most similar to the analytic model of Bate & Bonnell (2005, hereafter BB), which was developed in conjunction with numerical calculations of a turbulent, fragmenting cloud. BB assume that all protostars start with mass 0.003 $M_O$ due to the opacity limit for fragmentation. The protostars gain mass at a constant accretion rate, for a time period which follows a declining exponential probability. The two models differ physically, because the BB model starts from identical seed protostars, not accreting gas, because the constant growth rate does not reflect core-environment structure, and because the termination of accretion is due to dynamical ejection, not gas dispersal. Nonetheless, the models are mathematically similar, and BB obtain a mass function with a modal mass and high-mass slope similar to those of the IMF.

The mass functions of Figures 5 and 8 resemble the IMF more closely than do those of BB because the low-mass and high-mass slopes are closer to those summarized by Kroupa (2002). The low-mass slope of unity arises because at low mass, the mass accretion rate is due primarily to isothermal core gas, and thus is constant as in equation (12). Then in equation (16) for times earlier than the mean waiting time, $\Phi \propto M_\star$. The negative high-mass slope is more nearly constant in this model than in the BB model, because at high mass the mass accretion rate increases with mass, as in equation (13), while in the BB model the mass accretion rate is constant.

## 5. Summary

The main points presented in this paper are:

1. Models are presented for core-environment systems with continuous density profiles. Each profile has a core component which is a singular isothermal sphere, and an environment component which is either uniform in 3D, or an isothermal layer, or an isothermal cylinder.

2. Regions with low environment density $n_E$ are identified with regions of "isolated" star formation having low extinction and widely spaced cores, while regions with high $n_E$ resemble regions of "clustered" star formation having high extinction and more crowded cores.

3. The mass available for cold spherical accretion in a free-fall time increases from isolated to clustered regions. For isolated regions, this mass is low, independent of $n_E$, and arises mostly from the core component. For clustered regions, this mass is high, increases with $n_E$, and arises mostly from the environment component.

4. The properties of mass available in a free-fall time suggest that isolated regions can produce low-mass stars but not massive stars, while clustered regions can produce both low-mass and massive stars.

5. The distribution of protostar masses is computed from the competition between accretion and dispersal of core-environment gas. Assuming that the accreting gas can be dispersed at any time, the duration of accretion has a "waiting-time" distribution.

6. The protostar mass function matches the shape and mode of the IMF for cores with typical temperature 10 K, "clustered" environment density $3 \times 10^4$ cm$^{-3}$, mean waiting time 0.04 Myr, and a distribution of core masses, with HM values from 0.44 to 1.4 M$_O$. Near the peak of the mass function, most of the protostar mass comes from the core component. In the high-mass tail, most of the protostar mass comes from the environment component.

7. The protostar mass function has a high-mass tail matching that of the IMF for environment densities typical of clustered regions, but this feature decreases sharply for isolated regions. This behavior reflects the decrease in mass available in a free-fall time from clustered to isolated regions. The gas density between cores may be a key discriminant between isolated and clustered star formation, and between formation of low-mass and massive stars.

8. Cores in low-density filamentary environments are least likely to produce massive protostars, since their environment gas is likely to be dispersed before it can reach the protostar, and since the protostar accumulation length exceeds the typical extent of a cluster where massive stars are found.

**Appendix. Core-environment systems with nonuniform environments**

This Appendix presents expressions for the density and mass of core-environment systems where the environment is more centrally concentrated, and has lower dimension, than the uniform medium discussed in the main text.

## A.1. Density as a function of radius

A stratified 2D medium may be more realistic than a uniform 3D medium. If clusters are formed in regions compressed by OB winds or H II regions (Blaauw 1964, Elmegreen & Lada 1977, Myers 2009), a useful model may be the self-gravitating isothermal layer. Then the environment density is

$$n_E = n_l = n_{l0}\left[\sec h\left(\frac{z}{a_l}\right)\right]^2 \tag{A1}$$

where the layer density $n_l$ has peak density $n_{l0}$ when the distance $z$ from the midplane is zero, and where the scale height $a_l$ is

$$a_l \equiv \frac{\sigma}{(2\pi G m n_{l0})^{1/2}} \tag{A2}$$

with velocity dispersion $\sigma$ (Spitzer 1942). Here it is assumed for simplicity that the velocity dispersion of the layer gas is the same as that of the core gas. In a more detailed model, it may be more realistic for the velocity dispersion of the environment gas to exceed that of the core gas.

Centrally concentrated "filaments" are widely observed as hosts of star-forming dense cores in molecular clouds and in their young embedded clusters. In some star-forming regions multiple filaments diverge from a central hub, giving the appearance of a filamentary layer (Schmid-Burgk 1967, Myers 2009). A useful filament model is the self-gravitating isothermal cylinder, with density

$$n_E = n_f = n_{f0}\left(1 + \frac{x^2 + y^2}{a_f^2}\right)^{-2} \tag{A3}$$

where the filament density $n_f$ has peak value $n_{f0}$ on the $z$-axis where $x=y=0$, and where the scale height is

$$a_f \equiv \frac{\sigma}{(2\pi G m n_{f0})^{1/2}} \tag{A4}$$

(Stodolkiewicz 1963, Ostriker 1964).

Equations (A1)-(A4) are used along with equations (1) and (2) to obtain the curves in Figure 1.

### A.2. Mass as a function of radius

The mass of a core-environment system within spherical radius $r$, $M(r)$, is denoted as $M_l(r)$ when the environment is an isothermal layer and as $M_f(r)$ when the environment is an isothermal filament. Their masses are approximated by computing the mass centered on $r=0$, where the spherically symmetric component is enclosed by a sphere of radius $r$ and where the cylindrically symmetric component is enclosed by a right circular cylinder of radius $r_c$ and length $2r_c$. Here $r_c/r=(2/3)^{1/3}$ so that the cylindrical and spherical volumes are equal.

Then $M_l(r)$ is obtained by integrating equation (A1), giving

$$M_l(r) = \frac{2\sigma^2 r}{G}\left[1 + (12)^{-1/3}\xi_l \tanh\xi_l\right] \tag{A5}$$

where

$$\xi_l \equiv \left(\frac{3}{2}\right)^{1/3} \frac{r}{a_l}. \tag{A6}$$

Similarly, $M_f(r)$ is obtained by integrating equation (A3), giving

$$M_f(r) = \frac{2\sigma^2 r}{G}\left[1 + \left(\frac{16}{3}\right)^{1/3}\left(1+\xi_f^{-2}\right)^{-1}\right] \tag{A7}$$

where

$$\xi_f \equiv \left(\frac{3}{2}\right)^{1/3} \frac{r}{a_f}. \tag{A8}$$

Equations (A5)-(A8) are used to compute the curves in Figure 2. Equations (A5) and (A7) have form similar to that of equation (4) for the core-environment mass when the environment is uniform. They reduce to the core component alone for small radius, where the enclosed mass increases as $r$. At large radius, the enclosed mass increases as $r^D$, where $D=3, 2,$ and 1 when the environment is respectively uniform, a layer, and a filament.

### A.3. Mass as a function of free-fall time

The mass $M$ of a core-environment system as a function of its free-fall time $t_f$ is obtained from the above expression for $M(r)$ and from equation (7) for $M$ as a function of $r$ and $t_f$. When the environment is an isothermal layer, $M$ and $t_f$ are each calculated as functions of $r$. The resulting relation $M(t_f)$ is shown in Figures 3 and 4, but it has no simple analytic expression.

When the environment is an isothermal filament, $r$ can be eliminated, giving the expression

$$M_f = M_{core,f} \theta_f \left[ \frac{A - \alpha + \sqrt{(A-\alpha)^2 + 4\alpha}}{2} \right]^{3/2}. \quad (A9)$$

In equation (A9), the core mass is similar to the core mass when the environment is uniform, as in equation (6),

$$M_{core,f} = \left(\frac{6}{\pi}\right)^{1/2} \frac{\sigma^3}{G^{3/2} m^{1/2} n_{f0}^{1/2}}. \quad (A10)$$

The dimensionless free-fall time is normalized to the free-fall time based on the peak filament density,

$$\theta_f \equiv \frac{t_f}{t_{f0}}, \quad (A11)$$

with

$$t_{f0}^2 \equiv \frac{3\pi}{32Gmn_{f0}}. \tag{A12}$$

In equation (A9),

$$A = 1 + 2\left(\frac{2}{3}\right)^{1/3}, \tag{A13}$$

and

$$\alpha \equiv 2\left(\frac{2}{3}\right)^{1/3} \theta_f^{-2}. \tag{A14}$$

Equations (A9)-(A14) are used to produce the curves of core-filament mass as a function of free-fall time in Figures 3 and 4.

**Acknowledgements** This paper has benefitted from the support and encouragement of Irwin Shapiro, and from helpful discussions with João Alves, Shantanu Basu, Ted Bergin, Leo Blitz, Ian Bonnell, Bruce Elmegreen, Neal Evans, Adam Frank, Gary Fuller, Rob Gutermuth, Lee Hartmann, Patrick Hennebelle, Helen Kirk, Ralf Klessen, Charlie Lada, Chris McKee, Tom Megeath, Stella Offner, Frank Shu, and Hans Zinnecker. Many useful discussions occurred at the conference "SFR@50" in July, 2009 in Sarteano, Italy. The referee, Fred Adams, contributed numerous insightful comments and suggestions.

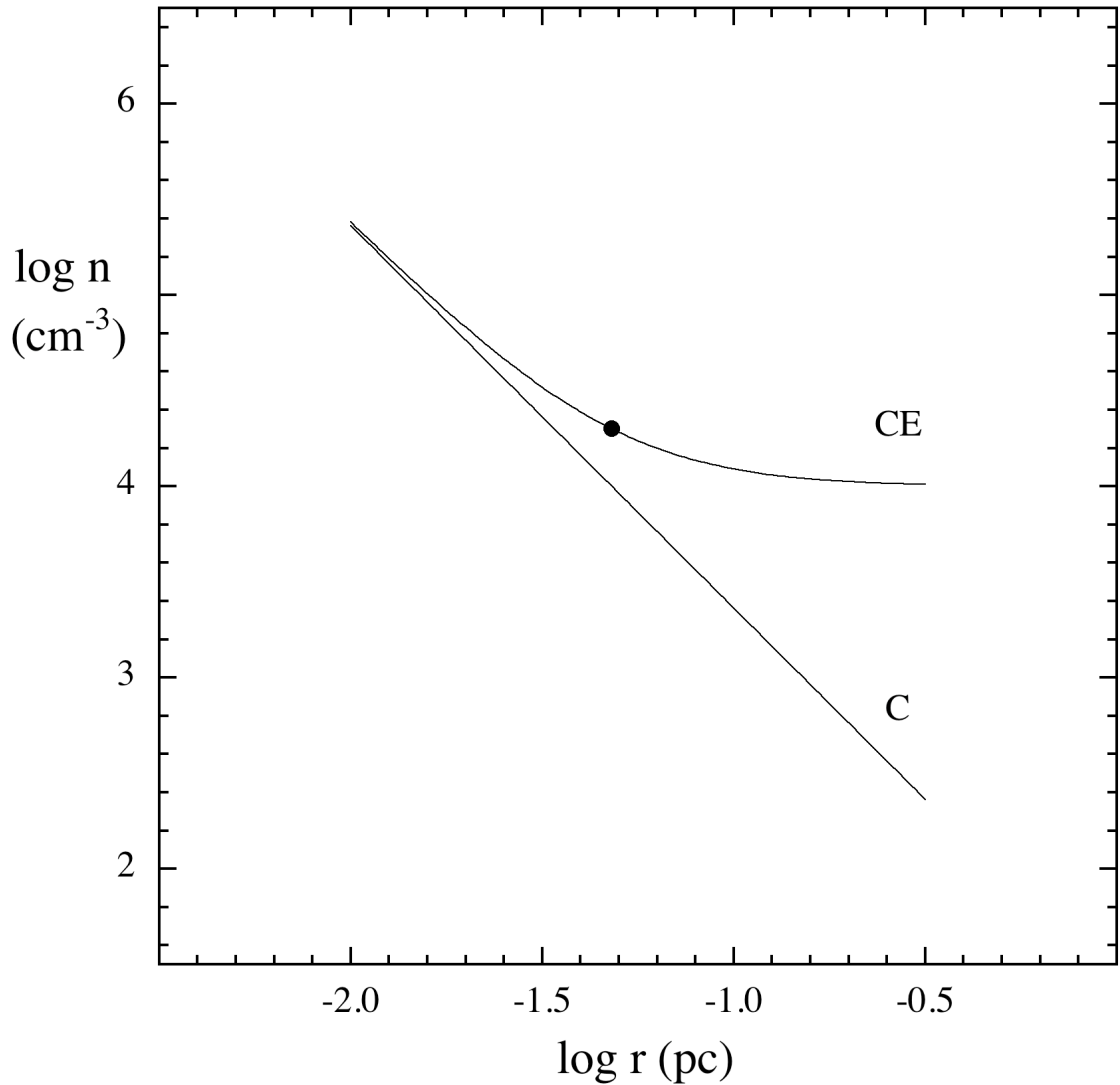

**Figure 1.** Density n of core-environment systems as a function of distance r from core center. Each environment model has peak density $n_E = 10^4$ cm$^{-3}$. The core profile *(C)* is for a singular isothermal sphere with temperature 10 K. The core-environment profile *(CE)* is for the system density along any direction when the environment is a uniform medium, along the midplane when the environment is an isothermal layer at 10 K, and along the symmetry axis when the environment is an isothermal filament at 10 K. The filled circle marks the equal-mass radius for a uniform environment, define in equation (5).

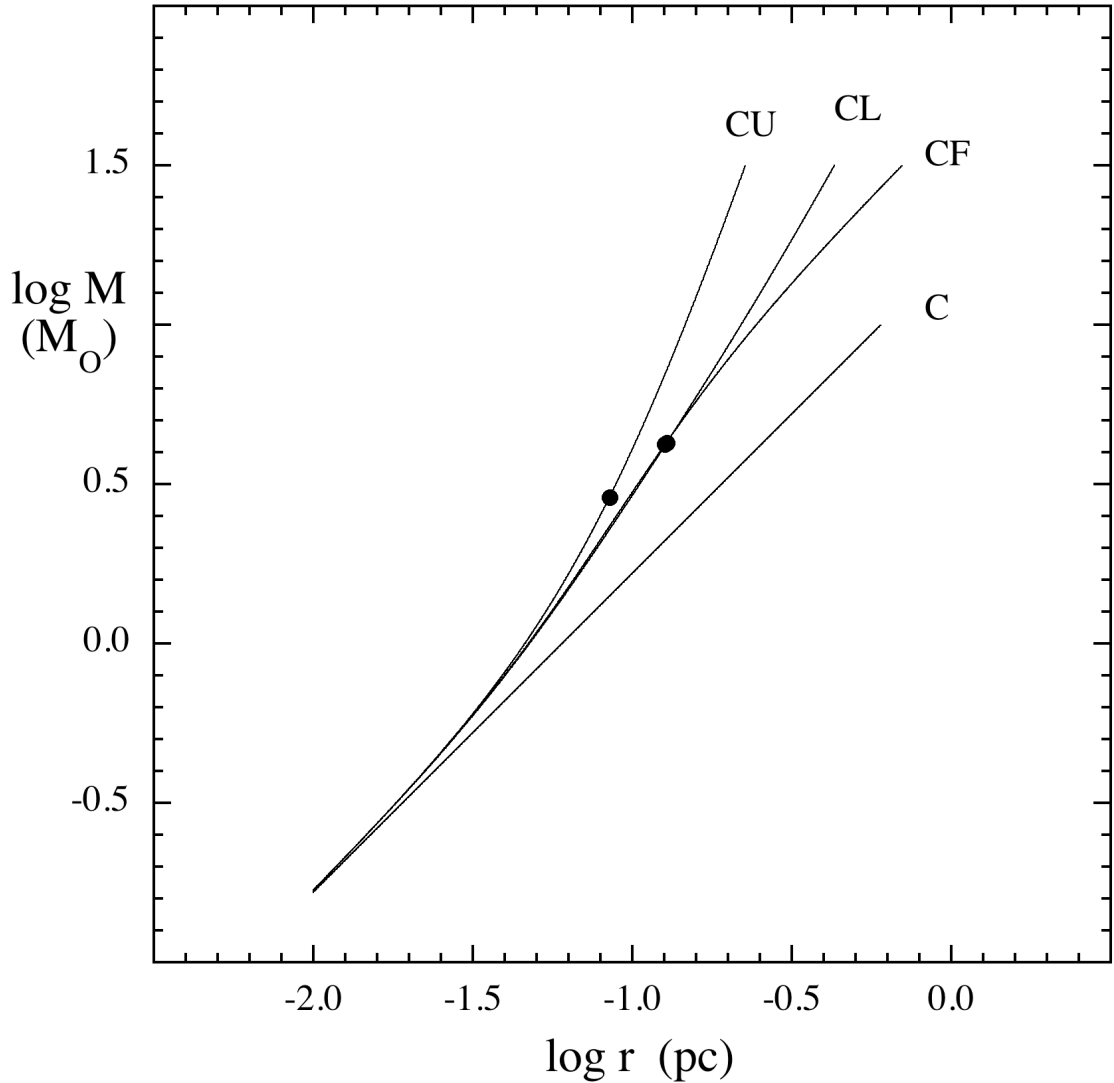

**Figure 2.** Mass $M$ within radius $r$, for the core and the core-environment systems whose density profiles are shown in Figure 1. The core $C$ is a singular isothermal sphere at 10 K. The core-environment systems have peak density $n_E = 10^4$ cm$^{-3}$. They are marked $CU$ for uniform environment, $CL$ for isothermal layer environment, and $CF$ for isothermal filament environment. Each filled circle marks the equal-mass radius as in Figure 1.

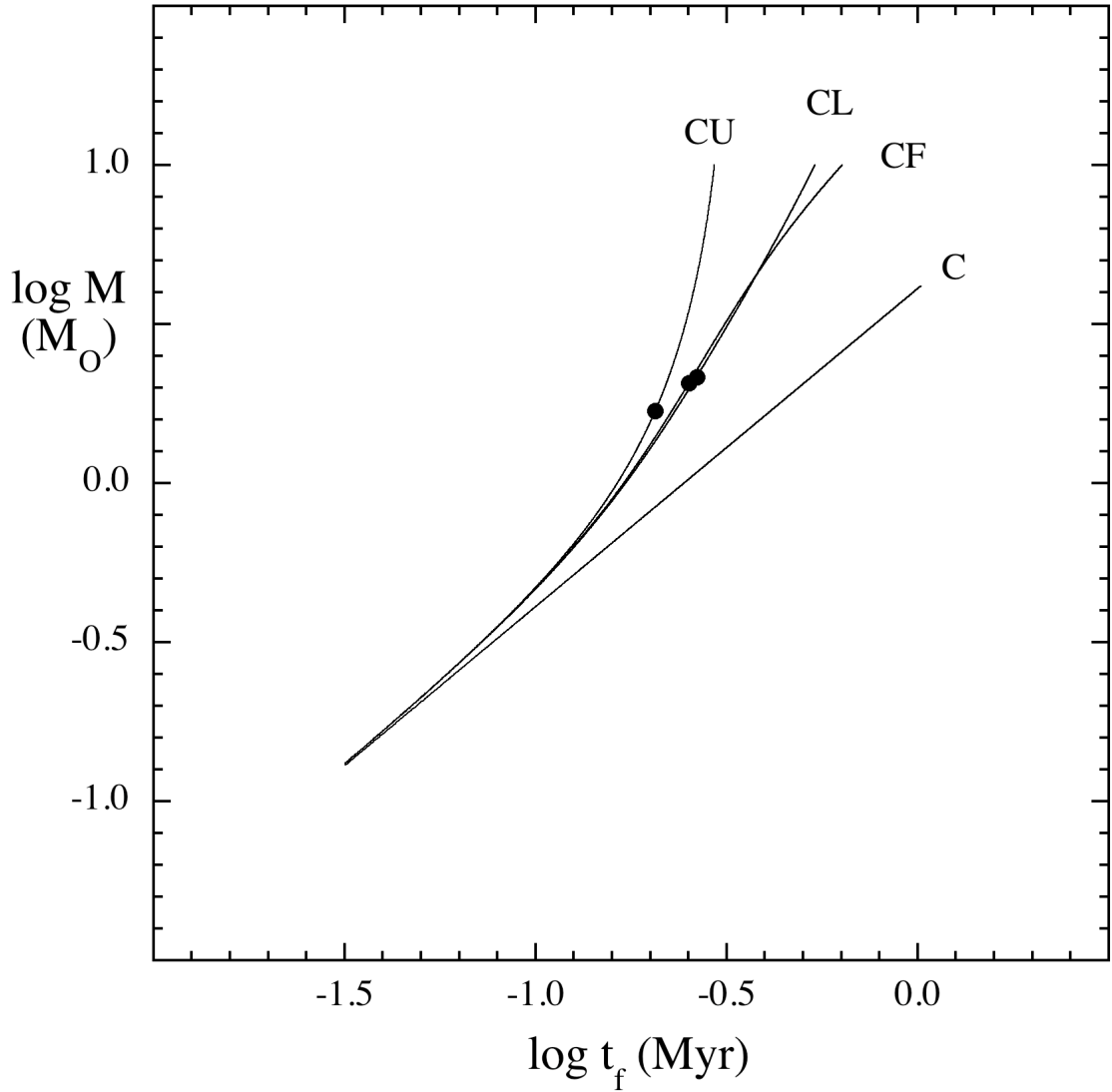

**Figure 3.** Gas mass *M* as a function of its free-fall time $t_f$, for the core and core-environment systems whose density profiles are shown in Figure 1. The core temperature is 10 K and the peak environment density is $10^4$ cm$^{-3}$. These curves are computed from equation (8) for the uniform environment, and from equations in the Appendix for environments which are isothermal layers and cylinders. Each filled circle marks the free-fall time at the equal-mass radius.

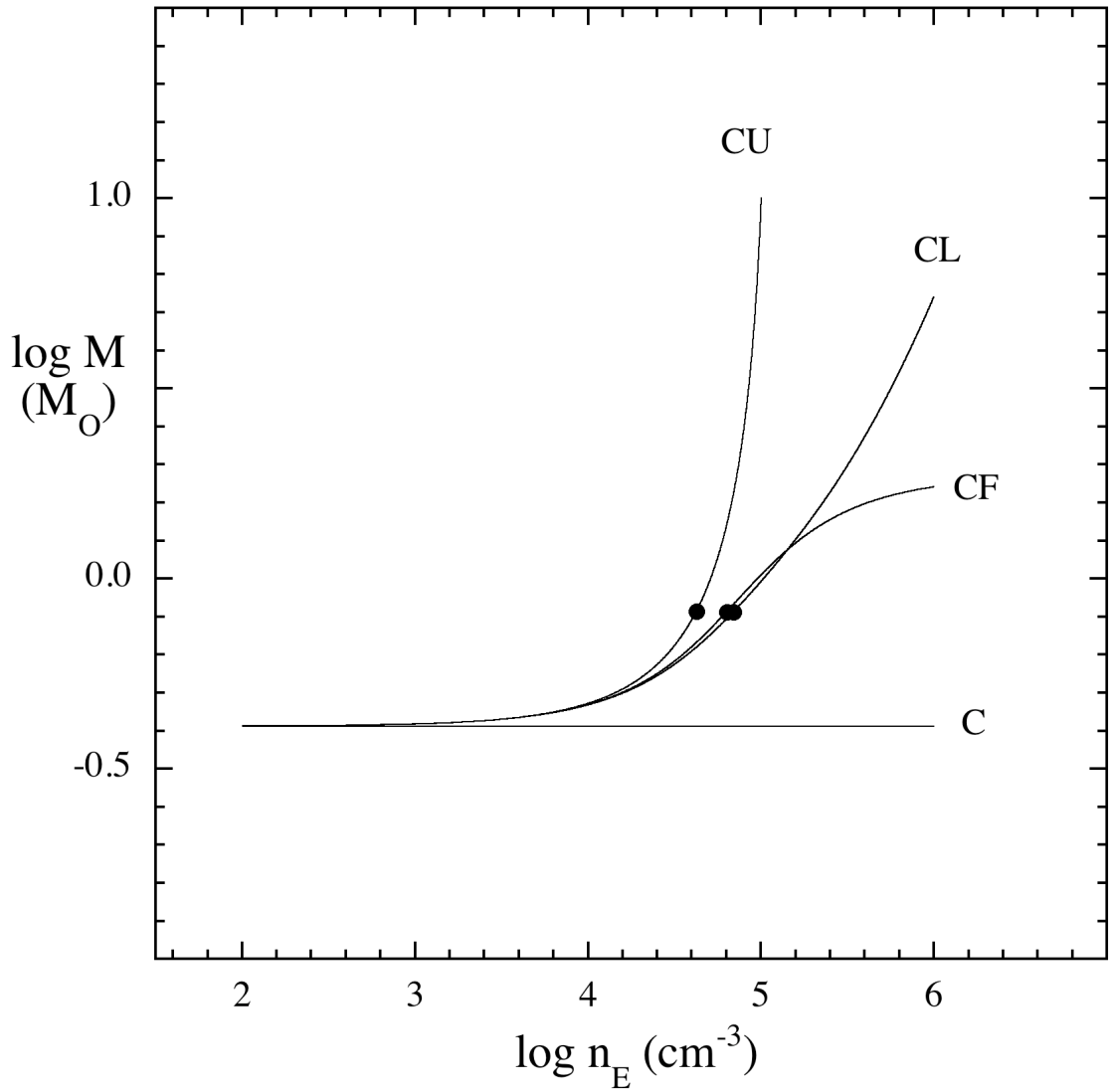

**Figure 4.** Mass of the core and core-environment systems in Figures 1-3, as a function of the peak environment density $n_E$, when the gas temperature is 10 K and the free-fall time $t_f$ is 0.1 Myr. Each filled circle marks the equal-mass environment density $n_{E0}$ defined in equation (10). Regions of low $n_E$ have low available mass, similar to that of low-mass stars, with most mass in the core component, independent of $n_E$. Regions of high $n_E$ have higher available mass, increasing with $n_E$, with most mass in the environment component. These two regimes resemble regions of "isolated" and "clustered" star formation.

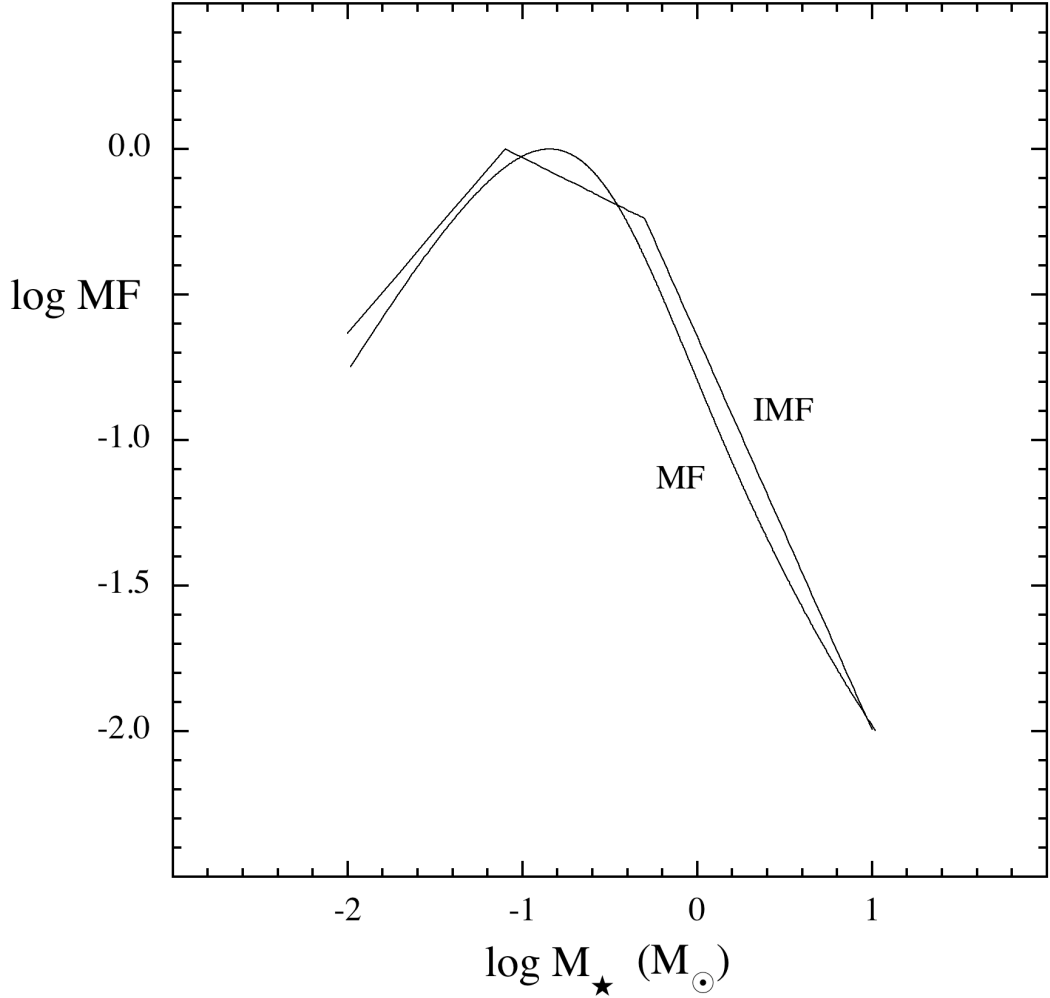

**Figure 5.** Mass function of protostars due to identical cores in a uniform environment, whose duration of accretion follows a waiting-time distribution. The shape is determined by the dimensionless mean waiting time $\bar{\theta}=0.20$, and the mode is determined by the product of the accretion efficiency and the core mass, $\varepsilon M_{core} = 0.80\ M_\odot$. The mass function *(MF)* resembles the initial mass function of stars *(IMF)* of Kroupa (2002) in its shape and position, but is slightly narrower. Masses of protostars near the peak arise mostly from the core component, while those in the high-mass tail arise mostly from the environment component.

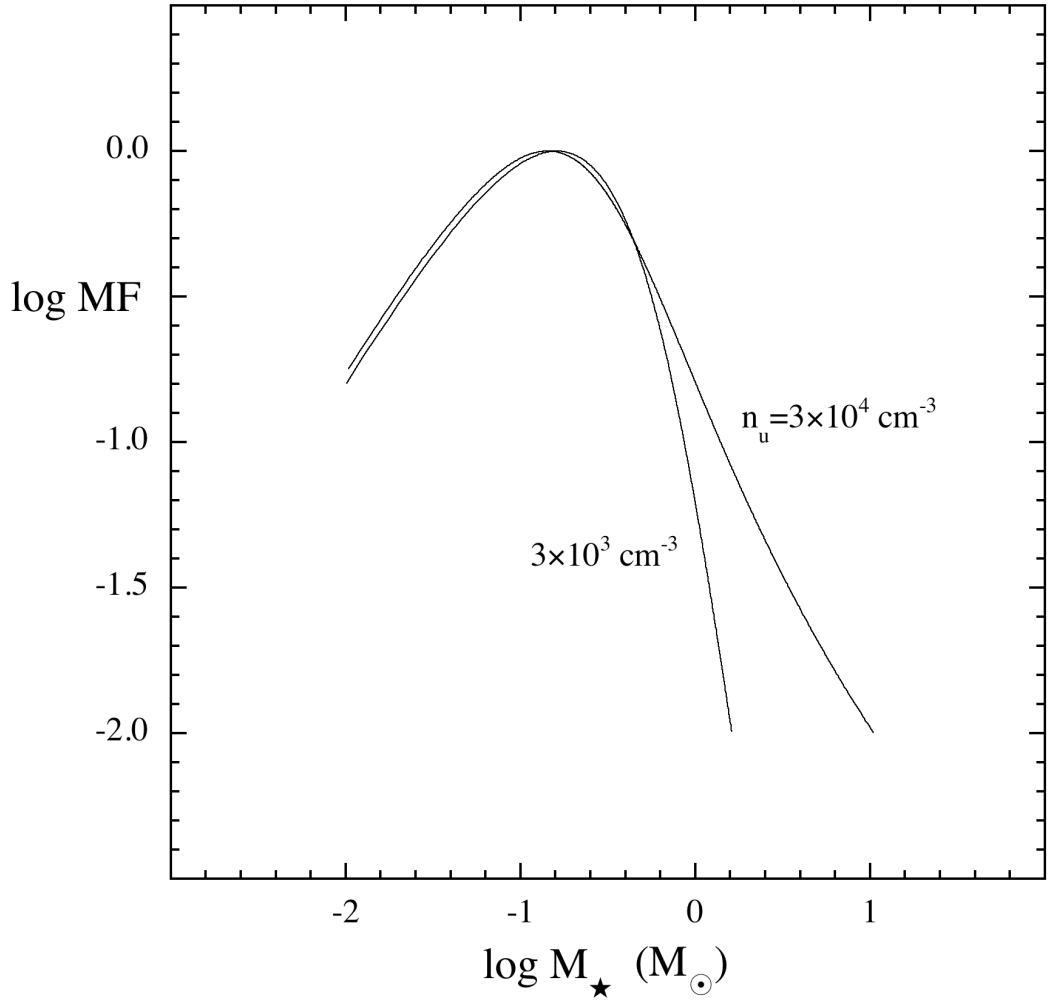

**Figure 6.** Mass functions of protostars based on temperature 10 K and mean waiting time 0.04 Myr, with environment gas density $3 \times 10^4$ cm$^{-3}$ as in embedded clusters, and $3 \times 10^3$ cm$^{-3}$ as in regions of isolated star formation. The high-mass tail seen for regions of high $n_E$ is absent for regions of low $n_E$. This suggests that clustered regions have a greater chance of producing massive stars than do isolated regions.

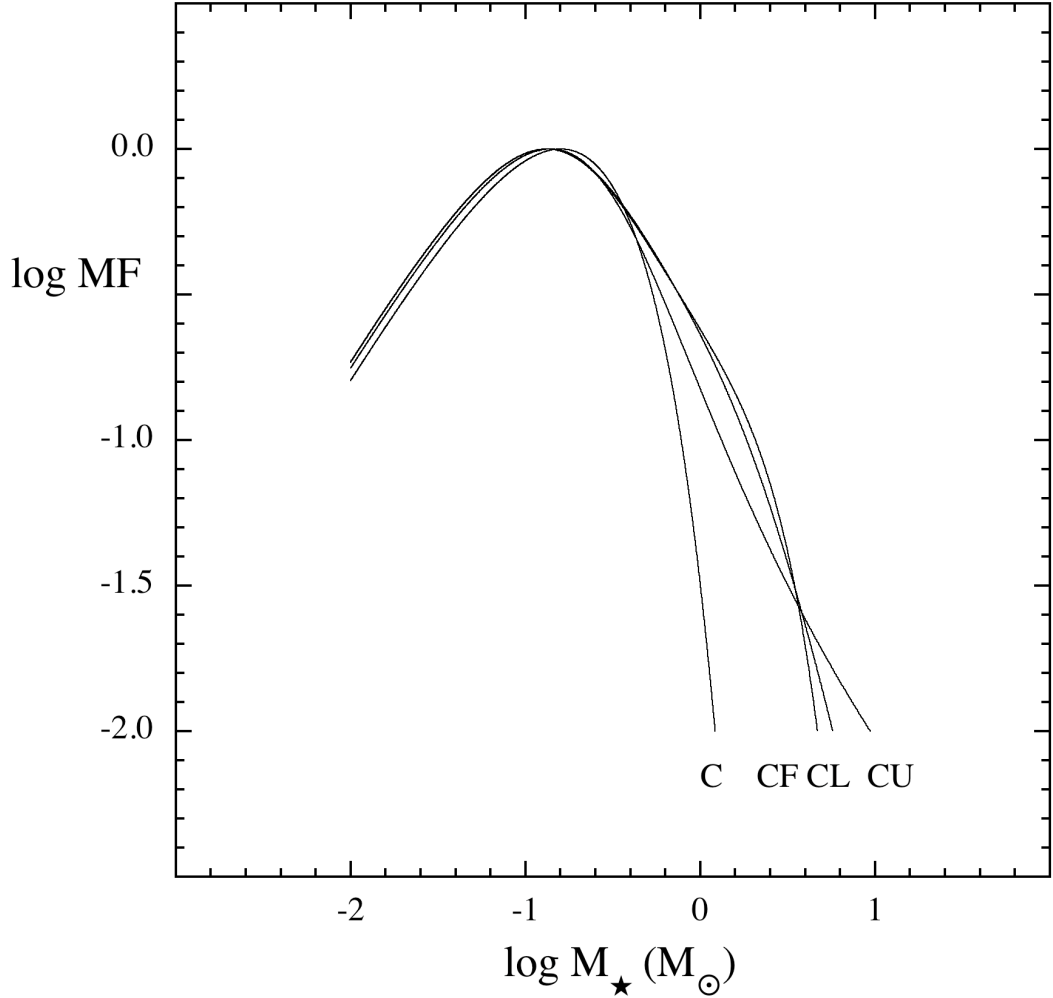

**Figure 7.** Mass function of protostars due to identical cores in various environments. The cores alone *(C)* represent the isolated limit of zero environment density. The environments in the systems *CF, CL,* and *CU* are respectively an isothermal filament, and isothermal layer, and a uniform medium. The accretion times follow a waiting-time distribution with mean waiting time 0.04 Myr. The temperature is 10 K, the peak environment density is $3 \times 10^4$ cm$^{-3}$, and the accretion efficiency is unity. All mass functions have nearly identical modes and low-mass slopes. The core-environment systems have slightly different high-mass tails, while the isolated cores have no high-mass tail.

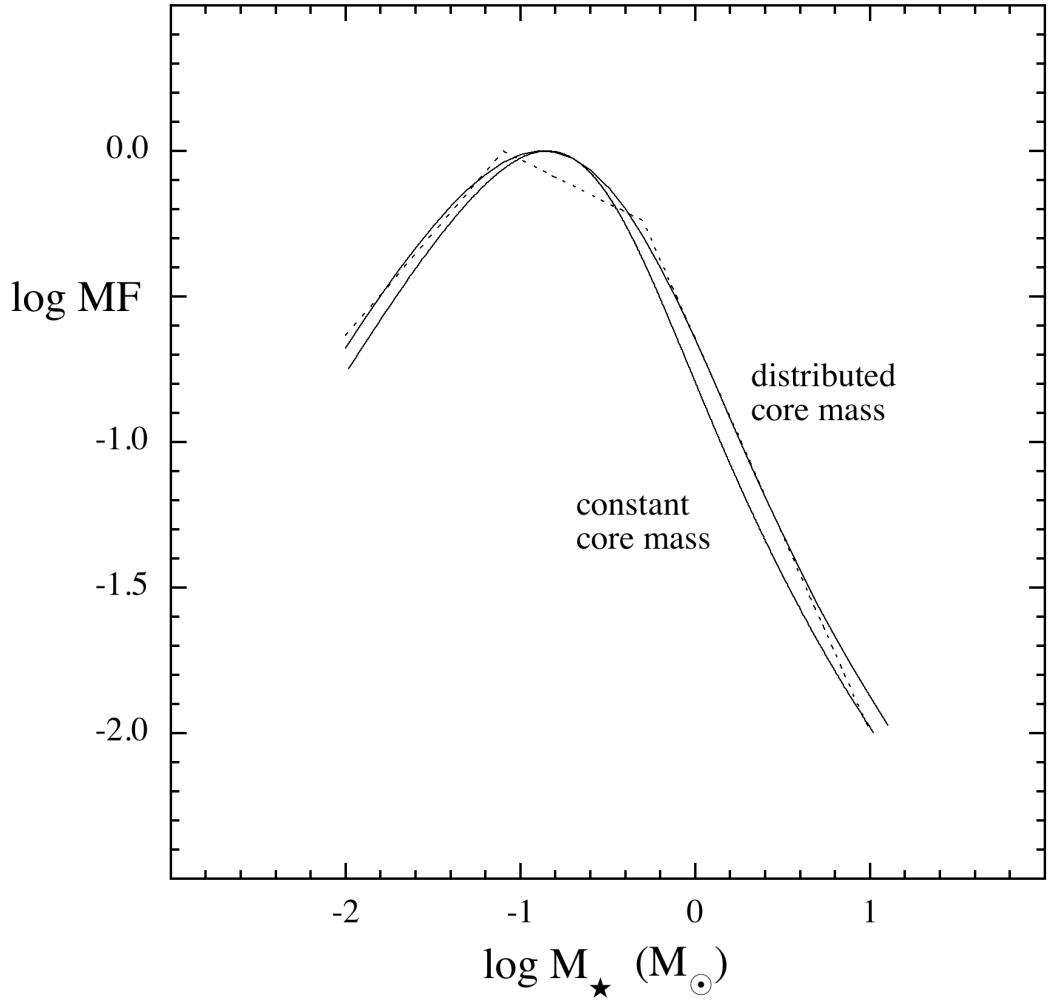

**Figure 8.** Protostar mass functions for core masses which are identical and distributed. Each mass function is based on the same environment density and mean waiting time as in Figures 5-7. The distributed masses follow a log-normal distribution whose mean $\mu$ corresponds to 0.8 $M_O$, or to typical temperature 10 K, the same as for the identical cores. The log-normal width parameter is $\sigma_{core}$=0.5, corresponding to half-maximum masses 0.44 and 1.4 $M_O$. This distribution of core masses makes a significant improvement in matching the IMF of Kroupa (2002), shown as a dotted line.